\documentclass[twoside,twocolumn,tightenlines, showpacs,aps]{revtex4}
\usepackage{graphicx}
\newcommand{\be}{\begin{equation}}
\newcommand{\ee}{\end{equation}}
\begin{document}

\title{Conductance peaks in open quantum dots}

\author{J. G. G. S. Ramos$^{1,2}$, D. Bazeia$^1$, M. S. Hussein$^2$, and C. H. Lewenkopf$^3$}
\affiliation{$^1$Departamento de F\'{\i}sica, Universidade Federal da Para\'{\i}ba, 58051-970 Jo\~{a}o Pessoa - PB, Brazil \\$^2$Instituto de F\'{i}sica, Universidade de S\~{a}o Paulo, 05314-970 S\~{a}o Paulo - SP, Brazil \\$^3$ Instituto de F\'{\i}sica, Universidade Federal Fluminense, 24210-346 Niter\'{o}i - RJ, Brazil}

\date{\today}

\begin{abstract}
We present a simple measure of the conductance fluctuations in open ballistic chaotic quantum dots, extending the number of maxima method originally proposed for the statistical analysis of compound nuclear reactions. The average number of extreme points (maxima and minima) in the  dimensionless conductance, $T$, as a function of an arbitrary external parameter $Z$, is directly related to the autocorrelation function of $T(Z)$. The parameter $Z$ can be associated to an applied gate voltage causing shape deformation in quantum dot, an external magnetic field, the Fermi energy, etc.. The average density of maxima is found to be $\left<\rho_{Z} \right> = \alpha_{Z}/Z_c$, where $\alpha_{Z}$ is a universal constant  and $Z_c$ is the conductance autocorrelation length, which is system specific. The analysis of $\left<\rho_{Z} \right>$ does not require large statistic samples, providing a quite amenable way to access information about parametric correlations, such as $Z_c$.
\end{abstract}


\maketitle

{\it Introduction} --
The statistical properties of the electronic transport in ballistic open quantum dots (QDs) have been intensively studied over the last decades \cite{beenakker97,alhassid00,mello}.
In such systems, the conductance can be described by the Landauer formula and, for QDs containing a large number of electrons, the random matrix theory (RMT) provides an excellent statistical description of the underlying chaotic electronic dynamics at the Fermi energy \cite{lewenkopf91,mello}.
RMT explains the observed universal conductance fluctuations in QDs, which depend only on the QD symmetries, such as time-reversal, and on the number of open modes $N$ connecting the QD to its source and drain reservoirs \cite{beenakker97}.

In the semiclassical limit of large $N$, the transmission statistical fluctuations are accurately modeled by Gaussian processes. In practice, it has been experimentally observed \cite{huibers98} and theoretically explained \cite{alves02} that, even for small values of $N$ and at very low temperatures, dephasing quickly brings the QD conductance fluctuations close to the Gaussian limit.

The conductance in open ballistic QDs exhibits random fluctuations as an external parameter, such as a magnetic field $B$ or an applied gate voltage $V_{\rm g}$, is varied. By identifying running averages with ensemble averages, it is customary to accumulate statistics  by varying as many parameters as the experimental set up allows. This invites one to ask whether useful statistical information can be extracted from the analysis of a single conductance curve. Inspired by the formal analogy between conductance and compound-nucleus Ericson fluctuations \cite{guhr98} we show that the answer is positive. More specifically, we further develop ideas originated
in the context of nuclear physics \cite{brink63}, to calculate the conductance average density of maxima and show its relation with the conductance autocorrelation function. As a result, we propose a new universal measure for the conductance of ballistic open QDs.

{\it Theoretical framework.} --
We consider the standard setting of a two-probe open quantum dot coupled by leads to a source and a drain electronic reservoirs. We also assume that the source (drain) reservoir is coupled to the quantum dot by a lead that has $N_1$ ($N_2$) open modes. The scattering matrix $S$ describing the electron flow is given by \cite{mello}
\begin{equation}
\label{eq:generalSmatrix}
 S = \left(
  \begin{array}{cc}
  r  & t\\
  t' & r'\\
  \end{array} \right)
\end{equation}
where $r$ $(r')$ is the $N_1 \times N_1$  ($N_2 \times N_2$) matrix containing the reflection amplitudes of scattering processes involving channels at the source (drain) coupled leads, while $t$ $(t')$ is the $N_1 \times N_2$ ($N_2 \times N_1$) matrix built by the transmission amplitudes connecting channels that belong to the source-coupled lead to the modes at the drain-coupled lead (and vice-versa).

At zero temperature, the linear conductance $G$ of an open quantum dot is given by the Landauer formula
\begin{equation}
\label{eq:landauer}
G = \frac{2e^2}{h} T \quad {\rm with} \quad T = {\rm tr} (t^\dagger t)
\end{equation}
where the factor 2 accounts for spin degeneracy and $T$ is the dimensionless conductance or transmission, which typically depends on $N_1$ and $N_2$, the quantum dot shape, the external magnetic field $B$, the electron energy $\varepsilon$, etc..

In the limit of large number of open modes, the average transmission for a chaotic QD is \cite{baranger94}
\begin{equation}
\langle T \rangle = \frac{N_1 N_2}{N_1+N_2}- \frac{\delta_{\beta,1}}{4},
\end{equation}
where  $\langle \cdots \rangle$ indicates that an ensemble average was taken and  $\beta=1$ ($\beta=2$) corresponds to the orthogonal (unitary) case of preserved (broken) time-reversal symmetry.  In the same  limit, the transmission correlation function reads \cite{efetov95,brouwer96,vallejos01}
\begin{equation}
\label{eq:correlationRMT}
\langle T^{\rm fl}(\epsilon, X)T^{\rm fl}(\epsilon',  X')\rangle
=\frac{\mbox{var}_\beta (T)} {[1 + (\delta X/X_c)^2)]^2 + (\delta \varepsilon/\Gamma)^2},
\end{equation}
where var$_{\beta}(T)=(1+\delta_{\beta,1})/16$. To simplify the notation, we introduce
$T^{\rm fl}(\varepsilon,  X) = T(\varepsilon,  X) - \langle T \rangle $, where $\varepsilon$ is the electron energy and $X$ is a generic parameter that describes a certain quantum dot shape belonging to a path of deformations caused by, for instance, applying a certain gate potential.
The correlation function given by Eq. \ref{eq:correlationRMT} is universal, with correlation length scales, $X_c$ and $\Gamma$, that are system dependent. There is a simple expression that relates $\Gamma$ to the mean resonance spacing $\Delta$, namely, $\Gamma=(N_1+N_2)\Delta/2\pi$ \cite{blatt52}.
The correlation length $\Gamma$ is generally different from the ``lifetime" or decay width, which is twice the imaginary part of the pole energy of the scattering matrix. Both quantities only coincide when
$\Gamma \ll \Delta$, a condition never met in {\it open} QDs.

 {\it Density of maxima.} --
The transmission or dimensionless conductance $T(Z)$ as a function of a generic parameter $Z$ (either $\varepsilon$ or $X$) has a maxima in the interval $[Z, Z+\delta Z]$ if
\begin{equation}
\label{eq:condition}
T^\prime(Z) > 0 \quad \mbox{and} \quad T^\prime(Z + \delta Z) < 0 \;,
\end{equation}
provided $\delta Z$ is small.  In this case, Eq.~\ref{eq:condition} implies that
\begin{equation}
\label{eq:integration-region}
- T^{\prime\prime}(Z) \delta Z > T^{\prime}(Z) > 0 \;.
\end{equation}
For convenience we introduce $T^\prime$ and $T^{\prime\prime}$ to denote the first and second derivatives of the dimensionless conductance $T$ with respect to $Z$.

The joint probability distribution $P(T^\prime,T^{\prime\prime})$ allows one to obtain the average density of maxima $\left<\rho_Z\right>$ \cite{brink63}: The probability to find a maximum in the interval $[Z,Z+\delta Z]$ is the integral of $P$ over the region defined in Eq. \ref{eq:integration-region}, that is
\begin{eqnarray}
\label{eq:construction}
\int_{-\infty}^0 dT^{\prime\prime} \int_0^{-T^{\prime\prime}\delta Z} dT^\prime
P(T^\prime,T^{\prime\prime}) = \nonumber\\
- \delta Z \int_{-\infty}^0 dT^{\prime\prime} T^{\prime\prime}
P(0,T^{\prime\prime}) \equiv \delta Z \left< \rho_Z \right >\,.
\end{eqnarray}

Let us infer $P(T^\prime,T^{\prime\prime})$ by examining the lowest moments of $T^\prime$ and $T^{\prime\prime}$.
Since the statistical properties of the dimensionless conductance are invariant under $Z$ translations, $T^\prime$ and $T^{\prime\prime}$ have zero mean. Their variance is directly related to the correlation function
\begin{eqnarray}
C_Z(\delta Z) = \left<T^{\rm fl}\!\left(Z + a \delta Z\right) T^{\rm fl}\!\left(Z - b \delta Z\right)\right> ,
\end{eqnarray}
which does not depend on the choice of $a$ and $b$, provided $a+b=1$. Neither do so the derivatives of $C_Z$ with respect to $\delta Z$, which leads to
\begin{eqnarray}
\left<[T^{\prime}]^2\right>& = - \left. \frac{d^2}{d (\delta Z)^2} C_Z(\delta Z) \right|_{\delta Z=0} \nonumber\\
\left<TT^{\prime\prime}\right>& = \left. \frac{d^2}{d (\delta Z)^2} C_Z(\delta Z) \right|_{\delta Z=0} \nonumber\\
\left<[T^{\prime\prime}]^2\right> & =  \left. \frac{d^4}{d(\delta Z)^4} C_Z(\delta Z) \right|_{\delta Z=0}\;,
\end{eqnarray}
and $\left<T T^\prime\right> = \left<T^\prime T^{\prime\prime}\right>=0$.
These results coincide and expand those obtained in Ref.~\onlinecite{rice54}.

We use the above relations and the maximum information principle to built the joint probability distribution of the transmission $T$ and its derivatives, $T^ \prime$ and $T^{\prime\prime}$. The distribution $P(T^\prime, T^{\prime\prime})$  is found by integrating over $T$, and gives
\begin{eqnarray}
P(0,T^{\prime\prime}) = \frac{1}{2\pi}\frac{1}{\sqrt{\left<[T^{\prime}]^2\right>\left<[T^{\prime\prime}]^2\right>}} \exp{\!\left(-\frac{1}{2}\frac{[T^{\prime\prime}]^2}{\left<[T^{\prime\prime}]^2\right>}\right)}.
\end{eqnarray}

Thus, the integral in Eq. \ref{eq:construction} renders
\begin{equation}
\left< \rho_Z \right> = \frac{1}{2\pi} \sqrt{\frac{\left<[T^{\prime\prime}]^2\right>}{\left<[T^{\prime}]^2\right>}}\;.
\end{equation}
This result, obtained with the help of the maximum information principle, is expected to be accurate in the large $N_1+N_2$ limit due to the central limit theorem \cite{note}. In the opposite limit of small $N_1+N_2$, the ratio $\left<T\right>/[\mbox{var}(T)]^{1/2}$ is no longer large and the constraint $ T  \ge 0$ gives raise to non trivial correlations between the transmission and its derivatives, as well, as deviations from the Gaussian distribution.

In the case where the external parameter $Z$ is the electron energy $\varepsilon$ and $N_1+N_2 \gg 1$, the correlation function $C_\varepsilon$ given by Eq.~\ref{eq:correlationRMT} reduces to a  Lorentzian
\begin{equation}
\label{eq:C-Lorentzian}
C_\varepsilon(\delta \varepsilon) =  \frac{\mbox{var}_\beta(T)}{1 +  \delta\varepsilon^2/\Gamma^2}.
\end{equation}
Such correlation function gives
\begin{equation}
\langle \rho_\varepsilon\rangle = \frac{\sqrt{3}}{\pi\Gamma} \approx \frac{0.55}{\Gamma} .\label{our1}
\end{equation}
Hence, by counting the average number of maxima in conductance one can infer the conductance correlation width. This idea was originally proposed as complementary to the analysis of the Ericson fluctuations in compound nucleus reactions \cite{brink63,bizzeti67,bonetti83}. The analysis of
Ref.~\onlinecite{bizzeti67}, seemingly different from ours \cite{foot}, gives the same result as above.

Support to our analytical findings is provided by numerical simulations employing the Hamiltonian approach to the statistical $S$-matrix \cite{vwz85}, namely
\begin{equation}
\label{eq:SHeidelberg}
S (\varepsilon) = \mbox{$\openone$} - 2\pi i W^\dagger (\varepsilon - H +
i \pi W W^\dagger)^{-1} W \;,
\end{equation}
where $\varepsilon$ is the electron propagation energy and $H$ is the matrix of dimension $M \times M$ that describes the resonant states. $H$ is taken as a member of the Gaussian orthogonal (unitary) ensemble for the (broken) time-reversal symmetric case. The matrix $W$ of dimension $M \times (N_1 + N_2)$ contains the channel-resonance coupling matrix elements.
Since the $H$ matrix is statistically invariant under orthogonal ($\beta = 1$) or unitary ($\beta = 2$) transformations, the statistical properties of $S$ depend only on the mean resonance spacing $\Delta$, determined by $H$, and $W^\dagger W$. We assume a perfect coupling between channels and resonances, which corresponds to maximizing the average transmission following a procedure described in Ref.\ \cite{vwz85}.
In this paper we restrict our numerical analysis to the $\beta=1$ case and, for simplicity, we take the case of $N\equiv N_1=N_2$.
We benchmarked the accuracy of the simulations by an extensive comparison between numerical simulations and analytical results \cite{baranger94} for  $\langle T \rangle$ and $\mbox{var}(T)$ as a function of $N$.

\begin{figure}[h!]
\begin{center}
\includegraphics[width=8.0cm,height=6.5cm]{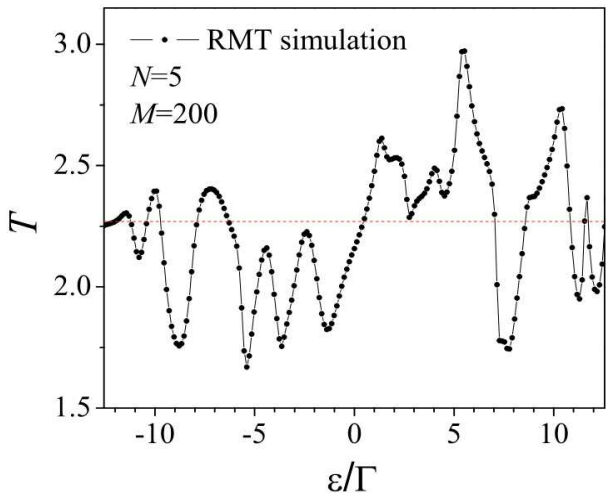}
\end{center}
\caption{Typical dimensionless conductance $T$ as a function of $\varepsilon$ for $N=5$ and perfectly transmitting modes (no direct processes). Black dots stand for the numerical results for a single realization of $H$ and the dotted line indicate the RMT prediction for $\langle T \rangle$.} \label{fig:TvsEN=5}
\end{figure}

Figure \ref{fig:TvsEN=5} illustrates the transmission $T(\varepsilon)$ for a typical realization of the matrix model given by Eq.\ \ref{eq:SHeidelberg}, for $N=5$ perfectly coupled modes close to the band center. The transmission correlation length is given by the Weisskopf estimate \cite{blatt52},
namely, $\Gamma=(N_1+N_2)\Delta/2\pi=N\Delta/\pi$.

\begin{figure}[h!]
\begin{center}
\includegraphics[width=8.0cm,height=6.5cm]{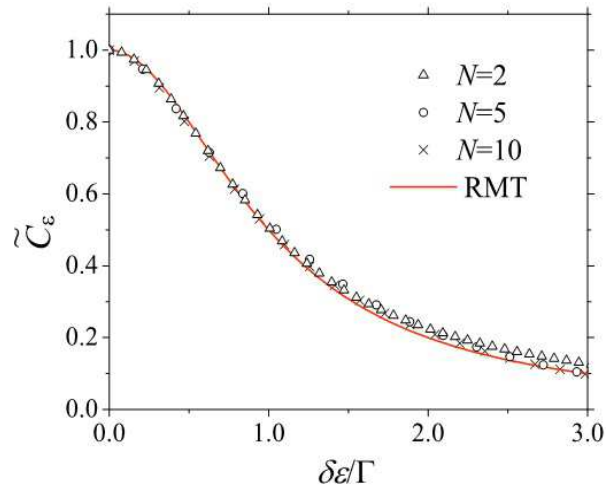}
\end{center}
\caption{Normalized transmission autocorrelation function $\widetilde{C}_\varepsilon(\delta \varepsilon) =	C_\varepsilon(\delta\varepsilon)/\mbox{var}(T)$ as a function of the energy $\delta \varepsilon$. Symbols correspond to ensemble averages for different number of channels $N$. The statistical error are smaller than the symbol sizes. The solid line stands for $\widetilde{C}_\varepsilon(\delta \varepsilon)$ given by Eq.\ \ref{eq:C-Lorentzian}.} \label{fig:CvsE-Nscaled}
\end{figure}

Figure \ref{fig:CvsE-Nscaled} shows the transmission autocorrelation function $C_\varepsilon(\delta \varepsilon)$ obtained from the model given by Eq.\ \ref{eq:SHeidelberg}, for $N$ perfectly coupled modes. The ensemble average is taken over 200 realizations of the $H$ matrices with $M=200$.
The random matrix theory \cite{vwz85} predicts an autocorrelation length $\Gamma = N \Delta/\pi$, which is nicely verified by the simulations.

\begin{figure}[h!]
\begin{center}
\includegraphics[width=8.0cm,height=6.5cm]{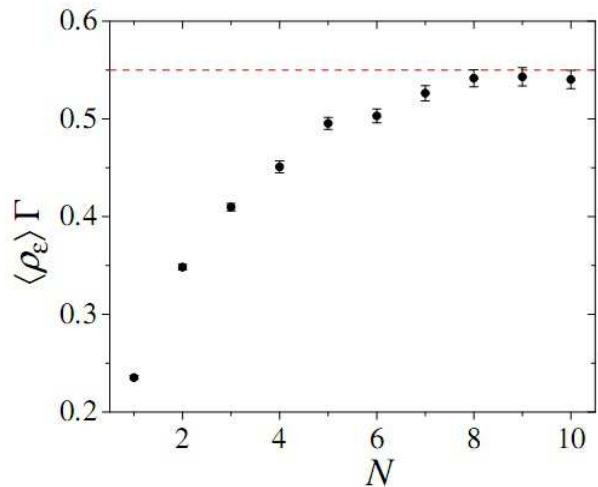}
\end{center}
\caption{Density of maxima $\langle \rho_\varepsilon \rangle \Gamma$ as a function of the number of open channels $N$. The symbols with statistical error bars correspond to our numerical simulations. The dashed line stands for the Gaussian process prediction.} \label{fig:rho-vs-N}
\end{figure}

Figure \ref{fig:rho-vs-N} shows the average density of maxima $\left< \rho_\varepsilon \right>$ in units of $1/\Gamma$ as a function of the number of open channels $N$. We observe that the agreement with the Gaussian process prediction becomes remarkably good as $N$ is increased.

Let us switch our analysis to the case where an external parameter modifies  the QD Hamiltonian, namely,  $H=H(X)$.
Taking $\delta \varepsilon =0$, the transmission autocorrelation function, Eq. \ref{eq:correlationRMT}, becomes a Lorentzian squared \cite{efetov95}
\begin{equation}
\label{eq:C-Lorentzian-sq}
C_X(\delta X) = \frac{\mbox{var}_\beta(T)}{[1 + (\delta X/X_c)^2]^2}.
\end{equation}
This correlation function gives for $\langle \rho_X\rangle$
\begin{equation}
\langle \rho_X\rangle = \frac{3}{\pi\sqrt{2} X_c} \approx \frac{0.68}{X_c}.\label{our2}
\end{equation}
The above result is new and is tested through numerical simulations in what follows.
\begin{figure}[h!]
\begin{center}
\includegraphics[width=8.0cm,height=6.5cm]{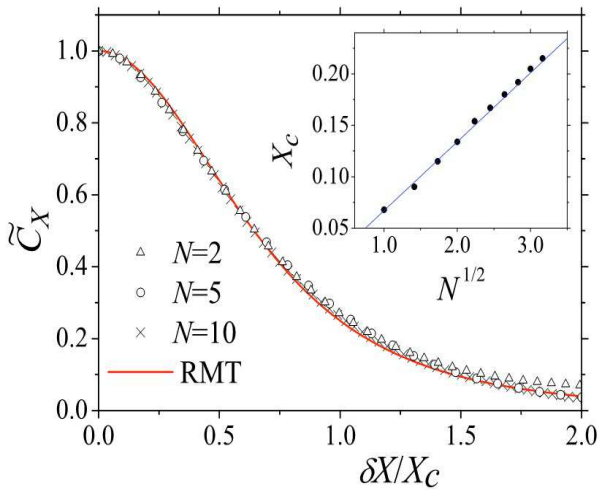}
\end{center}
\caption{Normalized transmission autocorrelation function
	$\widetilde{C}_X(\delta  X)=C_X(\delta X)/\mbox{var}(T)$ as a function of the parameter $\delta X/X_c$. Symbols
	correspond to numerical simulations for different $N$. The statistical error bars are smaller than symbol sizes. Solid line is
	given by theory, Eq.\ \ref{eq:C-Lorentzian-sq}.
  Insert:  $X_c$ versus $N^{1/2}$ showing a linear behavior, as indicated by the solid line.} \label{fig:CvsXN}
\end{figure}

To statistically model $H(X)$ we take $H=H_1\cos X + H_2 \sin X$ \cite{austin92}, where both $H_1$ and $H_2$ belong to a Gaussian ensemble. The transmission $T(X)$ is obtained by computing the $S$ matrix defined by Eq.\ \ref{eq:SHeidelberg} at $\varepsilon=0$  for 1,000 realizations of the $H(X)$ with  $M=200$. Figure \ref{fig:CvsXN} shows that a Lorentzian squared adjusts very nicely the numerically obtained correlation functions upon rescaling $X$ by $X_c$.

In distinction to the previous case, where a simple analytical expression for $\Gamma$ is known, here we determine $X_c$ numerically. Using semiclassical arguments, it can be shown \cite{baranger93} that the effect of a perturbation grows diffusively with the electron dwell time $t_{\rm D}$ in the quantum dot, which scales as $t_{\rm D} \sim 1/N$. Hence $X_c \sim \sqrt{N}$, in excellent agreement with our numerical findings, shown in the inset of Fig.~\ref{fig:CvsXN}.

\begin{figure}[h!]
\begin{center}
\includegraphics[width=8.0cm,height=6.5cm]{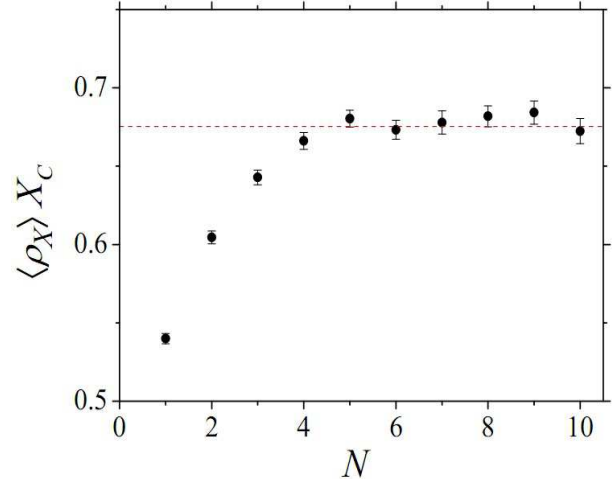}
\end{center}
\caption{Density of maxima $\langle \rho_X \rangle$ as a function of the number of open channels $N$ in units of $X_c$. The symbols with statistical error bars correspond to our numerical simulations. The dahed line stands for the theoretical prediction, Eq.~\ref{our2}.} \label{fig:rhoX-vs-N}
\end{figure}

Figure \ref{fig:rhoX-vs-N} summarizes our numerical results for the case of parametric Hamiltonian changes: The density $\langle \rho_X \rangle$ increases with $N$ and rapidly saturates at a value in very good agreement with our Gaussian process prediction given by Eq.~(\ref{our2}).
{\it Conclusions}-- In this work we extended the number of maxima method, originally employed in compound nuclear reactions, to open chaotic QDs. We have shown that the average density of maxima in the dimensionless conductance is inversely proportional to its autocorrelation length. For parametric variations that give rise to a Lorentzian-like transmission autocorrelation function, like variations in the electron energy, the universal proportionality constant is $\sqrt{3}/ \pi$. For parametric changes that lead to squared Lorentzian-like transmission correlations, such as gate potential variations, the universal proportionality factor is $3/( \pi \sqrt{2}) $. These results are obtained by assuming that the transmission derivatives are Gaussian distributed, which is expected to be rather accurate  in the semiclassical limit of large $N$. We
employ numerical simulations to infer the precision of our results for an arbitrary $N$. We show that even for moderate values of $N$ the semiclassical prediction gives already qualitatively good results, within about 10\% precision.
Our results may prove useful for the analysis of measurements of the transmission in chaotic quantum dots: By counting the maxima of a simple magnetoconductance trace, it is possible to estimate with a rather good precision the dimensionless autocorrelation function. More generally, ballistic mesoscopic systems (and potentially diffusive ones)  showing conductance fluctuations, such as graphene flakes \cite{ujiie09,tikhonenko09,ojeda10}, are also potentially amenable to this analysis.

	This work is supported in part by the Brazilian funding agencies CAPES, CNPq, FAPESP, and the Instituto Nacional de Ci\^{e}ncia e Tecnologia de Informa\c{c}\~{a}o Qu\^{a}ntica-MCT.


\begin{thebibliography}{99}
\bibitem{foot}
Our analysis is somewhat different from that of Ref.~\onlinecite{bizzeti67}. The latter starts with $T(Z - \delta Z) < T(Z + \delta Z)  < T(Z)$ as a maximum condition for $T(Z)$, and considers $T(Z - \delta Z) , T(Z),$ and $T(Z + \delta Z) $ as zero mean Gaussian variables.  Their joint probability distribution is used to obtain
%
\[\langle \rho_Z\rangle = \frac{1}{\pi \delta Z}\tan^{-1}\sqrt{\left[4\frac{ C_{Z}(0) - C_{Z}(\delta Z)}{C_{Z}(0) - C_{Z}(2 \delta Z)}\right] - 1}\]
%
which, in the limit $\delta Z \ll Z$, coincides with our results.
\end{thebibliography}

\end{document}